# Astro Data Lab Spectral Viewer Requirements Document

Data Lab Document

Prepared by: L. Fulmer, S. Juneau, and C. Merrill

With contributions by: A. Bolton, D. Nidever, R. Nikutta, S. Ridgway, K. Olsen, and B. Weaver

# Signatures

| Approvers | Approvers |
|---|---|
| | |
| Catherine Merrill<br>CSDC Project Manager | Leah Fulmer<br>Co-Author |
| | |
| Stephanie Juneau<br>Data Lab Science Team Lead | Enter Signer’s Name (N/A for none)<br>Enter Signer’s Title |
| | |
| Enter Signer’s Name (N/A for none)<br>Enter Signer’s Title | Enter Signer’s Name (N/A for none)<br>Enter Signer’s Title |
| | |
| Enter Signer’s Name (N/A for none)<br>Enter Signer’s Title | Enter Signer’s Name (N/A for none)<br>Enter Signer’s Title |
| | |
| Enter Signer’s Name (N/A for none)<br>Enter Signer’s Title | Enter Signer’s Name (N/A for none)<br>Enter Signer’s Title |
| | |
| Enter Signer’s Name (N/A for none)<br>Enter Signer’s Title | Enter Signer’s Name (N/A for none)<br>Enter Signer’s Title |

# Revision Log

| Version | Date mm/dd/yyyy | Affected Section(s) | Engineering Change # | Reason/Initiation/Remarks |
|---|---|---|---|---|
| A | 08/17/2018 | ALL | None | Initial Release |
| A1 | 04/08/2019 | 5,6 | | Update to include MVP and time phase information |
| A4 | 04/15/2020 | ALL | None | Update logos and institution name |
| A5 | 02/13/2023 | Cover | None | Update list of contributors |
| | | | | |
| | | | | |
| | | | | |
| | | | | |
| | | | | |

## List of Figures

**No table of figures entries found.**

## List of Tables

# Abbreviations and Acronyms

| Phrase/Acronym | Description |
|---|---|
| AD | Applicable Document |
| BOSS | Baryon Oscillation Spectroscopic Survey |
| DESI | Dark Energy Spectroscopic Instrument |
| DL | Data Lab |
| eBOSS | Extended BOSS |
| FITS | Flexible Image Transport System |
| GUI | Graphical User Interface |
| MVP | Minimum Viable Product |
| NSF | National Science Foundation |
| NOIRLab | National Optical-Infrared Astronomy Research Laboratory |
| SDSS | Sloan Digital Sky Survey |
| VOSpace | Virtual Observatory Space (Distributed storage) |

**Table 1: Abbreviations and Acronyms**

# 1 Reference Documents and Drawings

The following documents of the exact issue shown form a part of this design document to the extent specified herein.

| Code | Document Number | Title or Description |
|---|---|---|
| AD1 | | none |
| AD2 | | |
| AD3 | | |
| AD4 | | |
| AD5 | | |

**Table 2: Applicable Documents**

| Location | Document Number | Title or Description |
|---|---|---|
| | | none |
| | | |
| | | |

**Table 3: Applicable Drawings**

# 2 Scope

## 2.1 Introduction

This document contains a description of the requirements for a spectral viewer tool to be incorporated within the Data Lab environment at NSF's NOIRLab. The spectral viewer will be needed in order for Data Lab users to be able to visually inspect spectra. For each object, the spectral viewer will display the observed spectrum and, if available, the noise spectrum, sky spectrum, and best-fit template spectrum. Users will be able to control the display interactively after they launch the tool as part of their Data Lab workflow. The primary objective will be to support the visualization of spectroscopic datasets hosted at the Data Lab.

The spectral viewer may be adapted from an openly-distributed software package and, augmented by the Data Lab team as needed in order to fulfill all the requirements.

## 2.2 Important Definitions

The primary discriminator in the use of “shall” and “will” in this document is the resulting verification. “Shall” indicates a specific verifiable requirement. “Will” is descriptive, a statement of intent, or used to identify responsibility, an expectation, or for clarification. Verifications are not required for “will” statements.

| Term | Definition |
|---|---|
| Shall | "Shall" denote requirements that are mandatory and will be the subject of specific acceptance testing and compliance verification. |
| Can, May, or Should | "Can", "May", or "Should" indicate recommendations and are not subject to any requirement acceptance testing or compliance verification by the supplier. The supplier is free to propose alternative solutions. |
| Is or Will | "Is" or "Will" indicate a statement of fact or provide information and are not subject to any requirement acceptance testing or verification compliance by the supplier. |

**Table 4. Requirement term definitions**

| Term | Definition |
|---|---|
| Analysis | Analysis is the use of established technical or mathematical models or simulations, algorithms, or other scientific principles and procedures to provide evidence that the item meets its stated requirements. Analysis (including simulation) is used where verifying to realistic conditions cannot be achieved or is not cost effective and when such means establish that the appropriate requirement, specification, or derived requirement is met by the proposed solution. |
| Inspection | Inspection is an examination of the item against applicable documentation to confirm compliance with requirements. Inspection is used to verify properties best determined by examination and observation (e.g., paint color, weight, size etc.). |
| Test | A test is an action by which the operability, supportability, or performance capability of an item is verified when subjected to controlled conditions that are real or simulated. These verifications often use special test equipment or instrumentation to obtain very accurate quantitative data for analysis. |
| Demonstration | Demonstration is the actual operation of an item to provide evidence that it accomplishes the required functions under specific scenarios. Given input values are entered and the resulting output values are compared against the expected output values. |

**Table 5. Requirement Verification term defintion**

# 3 Data Lab Requirements

The Data Lab is preparing to host spectroscopic datasets. In particular, it will host a copy of the Dark Energy Spectroscopic Instrument survey, which is projected to include over 40 million spectra of galaxies and quasars as well as over 10 million spectra of stars. In addition, it will host a copy of the spectra from the Sloan Digital Sky Survey (SDSS), including Baryon Oscillation Spectroscopic Survey (BOSS), and Extended BOSS (eBOSS) spectra.

A key functionality that will be added to Data Lab will consist of a spectral viewer tool in order to visually inspect DESI, and SDSS/BOSS spectra. Given the large size of these spectroscopic datasets, a typical use case will consist of a selection or query for a subset of objects of interest (e.g., a subsample of stars or galaxies or quasars), followed by visual inspection of the selected spectra. It is anticipated that in some cases, users will want to go through a long list of spectra (e.g., thousands) quickly while looking for specific features by eye. In order to support this use case, the spectral viewer will need to load a list of objects, and quickly update the display from one object to the next on command from the user.

Overall, the spectral viewer tool should be integrated within the Data Lab environment, take as input a list that is a direct (or converted) result of running a query against Data Lab database spectroscopic tables, retrieve and display the spectra of interest quickly, and save a possible output within the Data Lab environment (e.g., user VOSpace).

# 4 Spectral Tool Requirements (L3)

## 4.1 DL_L3_SPEC_1001 Display Speed

We want the tool to be able to load a large set of spectra (e.g. thousands) and then display consecutive spectra quickly for rapid user evaluation. Use case: Looking for particular features within a group of thousands of spectra.

### 4.1.1 DL_L3_SPEC_1001.01

The spectral tool shall navigate from one spectrum to the next (or previous) through either clicking a button on the GUI or submitting a command within a Jupyter cell.

Verification: Inspection

Expected Evidence: In this case the inspection will include a video screen capture of toggling to next/previous spectra by button click and cell execution.

### 4.1.2 DL_L3_SPEC_1001.02

For retrieved spectra, the visualization tool shall display plots in a continuous update of less than or equal to 0.75 sec.

Verification: Test

Expected Evidence: The verification by test will include a minimum of 5000 cached spectra, averaged time report for 50 instances of spectral display.

### 4.1.3 DL_L3_SPEC_1001.03

The spectral tool shall maintain a cache of queried spectra that will display individual spectra according to the requirement DL_L3_SPEC_1001.02.

Verification: Test

Expected Evidence: For a minimum of 10000 queried spectra, averaged time report for 10 instances of loading --> caching. The development team will determine the correct caching scheme and prove its stability and performance.

## 4.2 DL_L3_SPEC_1002 Use within a Jupyter Notebook/ JupyterLab Environment

Ideally, one could run a command within a Jupyter Notebook cell that would create a pop-up window within the JupyterLab environment (perhaps appearing like a desktop tool). The window could be controlled simultaneously through scripting within the Jupyter Notebook as well as through the GUI. Use case: Sharing one's query / analysis with collaborators by sending an .ipynb file. Regarding scripting / GUI controls: accommodating different user approaches / physical abilities

### 4.2.1 DL_L3_SPEC_1002.01

The spectral tools shall operate within the Jupyter notebook OR JupyterLab environments.

Verification: Inspection

Expected Evidence Video screen capture of user controlling the spectral tool within a Jupyter Notebook (minimum number of callable tasks / "script" for these videos)

### 4.2.2 DL_L3_SPEC_1002.02

The spectral tool shall be controllable via commands within a Jupyter notebook cell or through a Python command line.

Verification: Inspection

Expected Evidence: Video screen capture of user controlling the spectral tool within the GUI

### 4.2.3 DL_L3_SPEC_1002.03

The spectral tool shall be controllable through a Data Lab GUI supplied interface within the Jupyter environment.

Verification: Inspection

Expected Evidence: Video capture of user controlling the spectral tool through the command line.

### 4.2.4 DL_L3_SPEC_1002.04

The spectral tool shall accept at least the data formats described by DESI and SDSS within
DESI:
https://desidatamodel.readthedocs.io/en/latest/DESI_SPECTRO_REDUX/SPECPROD/spectra-NSIDE/PIXGROUP/PIXNUM/spectra-NSIDE-PIXNUM.html

SDSS:
https://data.sdss.org/datamodel/files/BOSS_SPECTRO_REDUX/RUN2D/PLATE4/spPlate.html.

Verification: Test

Expected Evidence: Success / error report for input of different file types.

## 4.3 DL_L3_SPEC_1003 Dynamic Visualization of Spectra

### 4.3.1 DL_L3_SPEC_1003.01

The spectral tool shall pan and zoom spectra on command, including a reset view button.

Verification: Inspection

Expected Evidence: Video screen capture of tool displaying panning and zooming capabilities.

Discussion: It will be important for the spectral tool to offer dynamic visualization so that users can inspect each part of the spectra in detail.

### 4.3.2 DL_L3_SPEC_1003.02

The spectral tool shall provide a zoom box automatically.

Verification: Test

Expected Evidence: Video screen capture of tool displaying zoom box capabilities.

### 4.3.3 DL_L3_SPEC_1003.03

The spectral tool should offer the ability to zoom in on several predefined spectral windows at once, in the form of multiple zoom boxes, on command.

Verification: NOT A REQUIREMENT

Expected Evidence: Video screen capture of selecting several spectral windows, displaying multiple zoom boxes containing the selected spectral windows, and panning through multiple spectra to view their respective features within the selected windows.

Discussion: For Milky Way science, one can then look simultaneously at spectral windows containing lines of a specific element, which helps to diagnose the reality of a detection. The feature could also be useful for e.g. looking at the Balmer emission lines in a galaxy spectrum.

### 4.3.4 DL_L3_SPEC_1003.04

The spectral tool shall overlay spectra signal components (1D spectra, sky, noise) on command.

Verification: Inspection

Expected Evidence: Video screen capture of tool dynamically switching from overlayed to individual display.

Discussion: It can be helpful to have these spectra overlaid on the same graph or displayed on separate, stacked panels. We will want to construct a feature that can switch easily between these two modes if we implement both. Use case: Overlay spectra, sky noise for easier alignment. Separate spectra, sky noise for greater image clarity.

### 4.3.5 DL_L3_SPEC_1003.05

The spectral tool shall display each spectral signal component (1D spectra, sky, noise) individually on command.

Verification: Inspection

Expected Evidence: Video screen capture of tool dynamically switching from overlayed to individual display.

### 4.3.6 DL_L3_SPEC_1003.06

The spectral tool shall display each spectrum, template, and lines in observed or rest-frame wavelengths on command.

Verification: Inspection

Expected Evidence: Video screen capture of tool switching between observed and rest-frame wavelength display.

### 4.3.7 DL_L3_SPEC_1003.07

The spectral tool shall reset plots to original (unzoomed, unedited) format on command.

Verification: Inspection

Expected Evidence: Video screen capture of zooming and panning around a particular spectrum, then executing a reset plots command that effectively returns the spectral display to its original format.

Discussion: It will be necessary to offer users the option to reset plots after they zoom in / out unintentionally.

## 4.4 DL_L3_SPEC_1004 Template Fitting & Quality Check

DESI will automatically fit a template to each spectra, so we want to offer users a measurement of the quality of the fit (template fitting data) and residuals, as well as an opportunity to change the template redshift. Down the line, we could offer users the choice to redo the template fit, using built-in template libraries. We might be able to call on the Astropy community for help on this. Use case: Measuring stellar radial velocities, redshifts.

### 4.4.1 DL_L3_SPEC_1004.01

The spectral tool shall overlay the DESI calculated redshift template (as defined by the DESI pipeline) on command.

Verification: Test

Expected Evidence: Image screen capture of DESI redshift template overlaid on spectrum and confirmation that the overlay is consistent with DESI expected values.

### 4.4.2 DL_L3_SPEC_1004.02

The spectral tool shall overlay a user-defined redshift among existing templates on command.

Verification: Test

Expected Evidence: Image screen capture of existing template with user-defined redshift overlaid on spectrum and confirmation that the overlay is consistent with expected values.

### 4.4.3 DL_L3_SPEC_1004.03

The spectral tool shall overlay a user-uploaded template on command.

Verification: Inspection

Expected Evidence: Video screen capture of user selection various redshift templates and subsequently observing changes in the overlaid template, the X^2 value, the fitness quality value, and the wavelength solution. Verify that all values are consistent with expected/calculated results.

### 4.4.4 DL_L3_SPEC_1004.04

The spectral tools shall calculate a X^2 or other measurement of template fit quality.

Verification: Inspection

Expected Evidence: Video screen capture of user selection various redshift templates and subsequently observing changes in the overlaid template, the X^2 value, the fitness quality value, and the wavelength solution. Verify that all values are consistent with expected/calculated results.

### 4.4.5 DL_L3_SPEC_1004.05

The spectral tool shall display the fitness quality value automatically.

Verification: Inspection

Expected Evidence: Video screen capture of user selection various redshift templates and subsequently observing changes in the overlaid template, the X^2 value, the fitness quality value, and the wavelength solution. Verify that all values are consistent with expected/calculated results.

### 4.4.6 DL_L3_SPEC_1004.06

The spectral tool shall display the residual on command.

Verification: Test

Expected Evidence: Image screen capture of the tool displaying the residual. Verify that all values are consistent with expected / calculated results.

## 4.5 DL_L3_SPEC_1005 Plot Locations for Emission/Absorption Lines

Displaying different line types in different colors, options to display specific subsets of all possible lines (e.g. only strong lines, only emission lines, etc.), loading a user-defined list of spectra with user-defined colors, implementing an “addlines” function from the Jupyter Notebook. Connect line displays to adjustable template-fitting feature (above). Use case: More easily identifying which lines are depicted in the observed features.

### 4.5.1 DL_L3_SPEC_1005.01

The spectral tool shall overlay expected locations of emission / absorption lines on command in all relevant display panels, if a redshift is available (assume redshift equals zero if none available).

Verification: Test

Expected Evidence: Video screen capture of dynamic display of emission / absorption lines on command and verify that all values are consistent with expected/calculated results.

### 4.5.2 DL_L3_SPEC_1005.02

The spectral tool shall contain a library of emission / absorption lines for user selection in the overlay process.

Verification: Test

Expected Evidence: Video screen capture of dynamic display of emission / absorption lines on command and verify that all values are consistent with expected/calculated results.

### 4.5.3 DL_L3_SPEC_1005.03

The user shall be able to specify emission / absorption lines for display on command.

Verification: Test

Expected Evidence: Submission of input file and image screen capture of displayed emission/absorption lines.

## 4.6 DL_L3_SPEC_1006 Contextual Viewer

Similar to GlueViz, connecting visualized spectra to photometric images / location on the sky. Let's first see what capabilities already exist within the Jupyter Lab environment. Use case: Automatically connecting a user's object to external images and catalogs for a more dynamic analysis.

### 4.6.1 DL_L3_SPEC_1006.01

The spectral tool shall display a photometric image of the object corresponding to the spectra displayed.

Verification: Inspection

Expected Evidence: Image screen capture of contextual viewer

### 4.6.2 DL_L3_SPEC_1006.02

The contextual viewer should anticipate further applications of contextual capabilities (e.g. displaying a spectral energy distribution, 2D spectrum with trace, etc).

Verification: Not a Requirement

Expected Evidence: We want the spectral tool's contextual capabilities to allow users to complete successive database queries surrounding the currently-displayed object.

### 4.6.3 DL_L3_SPEC_1006.03

The spectral tool should give users the option to search around the displayed object within all Data Lab holdings.

Verification: Not a Requirement

Expected Evidence: We want the spectral tool's contextual capabilities to allow users to complete successive database queries surrounding the currently-displayed object.

## 4.7 DL_L3_SPEC_1007 Data Access and Saving

Allow users to access data through both the Data Lab query manager and by uploading files from their computer. Similarly, allow users to download FITS files of the displayed data / layers as well as saving data within the Data Lab VOSpace.

### 4.7.1 DL_L3_SPEC_1007.01

The spectral tool shall support default units of wavelength and flux.

Verification: Inspection

Expected Evidence: Example of input data with no units defined, processed and displayed with correct default units.

### 4.7.2 DL_L3_SPEC_1007.02

The spectral tool should support user-defined units for the x and y axes.

Verification: Not a Requirement

Expected Evidence: Example of input data with units pre-defined, processed and displayed with correct pre-defined units.

### 4.7.3 DL_L3_SPEC_1007.03

The spectral tool shall allow users to access spectral data through a Data Lab file access service / frontend.

Verification: Inspection

Expected Evidence: Success / error report from tailored query.

### 4.7.4 DL_L3_SPEC_1007.04

The Data Lab file access service / frontend shall query a spectroscopic table within a Data Lab database (e.g. queries according to RA, dec, exposure information, target class, etc.).

Verification: Inspection

Expected Evidence: Success / error report from tailored query.

### 4.7.5 DL_L3_SPEC_1007.05

The spectral tool shall handle authentication protocols provided by the Data Lab file access service / frontend, as needed to respect proprietary data access.

Verification: Test

Expected Evidence: Log of spectral tool processing and responding appropriately to both authorized and non-authorized accounts, allowing for data access on authorized accounts and an error message on non-authorized accounts.

### 4.7.6 DL_L3_SPEC_1007.06

The Data Lab file access service / frontend shall cache the queried spectra so that they are displayed according to the display requirements.

Verification: Test

Expected Evidence: Success / error report from caching attempt and video screen capture of query-to-display workflow.

### 4.7.7 DL_L3_SPEC_1007.07

The spectral tool should allow users to access spectral data by uploading files from their local computer.

Verification: Not a Requirement

Expected Evidence: Success / error report from file upload.

### 4.7.8 DL_L3_SPEC_1007.08

The spectral tool shall allow users to download FITS files of the displayed spectrum, including a template on command.

Verification: Test

Expected Evidence: The spectral tool shall allow users to download FITS files of the displayed spectrum, including a template on command.

### 4.7.9 DL_L3_SPEC_1007.09

The spectral tool shall allow users to save the displayed spectrum and template within the Data Lab VOSpace on command.

Verification: Test

Expected Evidence: Success / error report for successful saving to VOSpace. Confirmation that the cache be called and re-displayed accurately.

#### 4.7.10 DL_L3_SPEC_1007.10

The spectral tool shall save plots (image files) on command.

Verification: Test

Expected Evidence: Success / error report that plots were saved as images with all information retained.

#### 4.7.11 DL_L3_SPEC_1007.11

The spectral analysis steps conducted within the spectral tool shall be logged, and the log shall be saved on command.

Verification: Test

Expected Evidence: Success / error report of successful logging of all steps, either in command window or output file.

#### 4.7.12 DL_L3_SPEC_1007.12

The spectral tool should save the final state of the tool following a session on command.

Verification: Not a Requirement

Expected Evidence: Success / error report from saved session, indicating that all information about the current session was successfully saved in output file.

### 4.8 DL_L3_SPEC_1008 Catalog Information and Annotations

This is included within the DESI Data Interface Visualization Requirements, and would be equally valuable for the Data Lab spectral visualization tool.

#### 4.8.1 DL_L3_SPEC_1008.01

Spectral tool shall display available catalog information such as spectroscopic classification and sub-class, redshift, redshift uncertainty, ZWARN flags, TARGETID and targeting bits, exposure number, fiber number, imaging data such as magnitudes, profiles, etc. automatically.

Verification: Inspection

Expected Evidence: Video screen capture of catalog information changing dynamically as user toggles through several spectra.

### 4.8.2 DL_L3_SPEC_1008.02

Upon saving a file, the spectral tool shall allow users to annotate the displayed spectra (with a flag or comment) and store annotations within the saved file on command.

Verification: Test

Expected Evidence: Success / error report from saved file (was the annotation saved as expected.

## 4.9 DL_L3_SPEC_1009 Temporal Analysis of Spectra

### 4.9.1 DL_L3_SPEC_1009.01

The spectral tool should provide information from individual exposures of a particular object by default.

Verification: Not a requirement

Expected Evidence: Video screen capture of dynamically switching from one display type to the other.

### 4.9.2 DL_L3_SPEC_1009.02

The spectral tool should provide information from individual exposures of a particular object on command.

Verification: Not a Requirement

Expected Evidence: Video screen capture of dynamically switching from one display type to the other.

## 4.10 DL_L3_SPEC_1010 Perform Functions on Spectra

It is important for visual inspection and comprehension that users be able to smooth a given spectra, as well as perform addition and subtraction in flux space on displayed spectra.

### 4.10.1 DL_L3_SPEC_1010.01

The spectral tool should perform spectral addition and subtraction in flux space on command.

Verification: Not a Requirement

Expected Evidence: Success / error report for a series of given functions (add, subtract, …)

### 4.10.2 DL_L3_SPEC_1010.02

The spectral tool shall display a smoothed spectrum using a user-specified smoothing level and common smoothing algorithms on command, including an option to return to the unsmoothed spectrum.
Verification: Test

Expected Evidence: Success / error report from smoothing attempt, given different smoothing input values.

# 5 Minimum Viable Product

## 5.1 Definition and Discussion

The Minimum Viable Product (MVP) is a concept that relates to the phased development approach for Data Lab and all of the CSDC products. The MVP is defined by the science team for each product as the core or minimum capabilities of a product in order to deliver the functions required of the tool. In the case of the Spectral Viewer, Data Lab identified a full set of requirements, as captured in Revision A of this document. In Rev B, the science team has evaluated all of the requirements and prioritized them. In essence, if all of the MVP requirements are not met – at some capacity – then the tool will not be considered for release as part of Data Lab.

The phased approach to development indicates a culture of continuous improvement of the product. However each release of the tools will be response to a) a bug fix which will cause the product to go from Version X.x to X.x+1, or b) in increase in the functional or system performance capabilities of the product. These changes are denoted as X.X to X+1.0.

In the case of the spectral viewer, the capabilities of several open-source spectral visualization tools are being evaluated for incorporation into Data Lab. As it is quite possible that no one tool will have all of the capabilities listed here, it is important to note which functions are the highest priority. Tools will be evaluated for compliance based on the full set. If the full set of requirements is not met, but the MVP requirements are met, Data Lab can incorporate the tool and then iterate development of the other capabilities as the rest of Data Lab matures. If none of the tools meet the MVP requirements however, a tool will be selected for maturation. Only once the tool has proven MVP capabilities will it be incorporated into Data Lab for release.

# 6 Requirements Verification Matrix

| Requirement number | Title or Heading | Text | Verification Method | | | | MVP? (y/n) |
|---|---|---|---|---|---|---|---|
| | | | A | T | I | D | |
| DL_L3_SPEC_1001 | Display speed | | | | | | |
| **DL_L3_SPEC_1001.01** | | The spectral tool shall navigate from one spectrum to the next (or previous) through either clicking a button on the GUI or submitting a command within a Jupyter cell. | | | X | | **Y** |
| **DL_L3_SPEC_1001.02** | | For retrieved spectra, the visualization tool shall display plots in a continuous update of less than or equal to .75 sec. | | X | | | **Y** |
| **DL_L3_SPEC_1001.03** | | The spectral tool shall maintain a cache of queried spectra that will display individual spectra according to the requirement DL_L3_SPEC_1001.02. | | X | | | **Y** |
| DL_L3_SPEC_1002 | Use within a Jupyter Notebook / JupyterLab environment | | | | | | |
| **DL_L3_SPEC_1002.01** | | The spectral tools shall operate within the Jupyter notebook OR JupyterLab environments. | | | X | | **Y** |
| **DL_L3_SPEC_1002.02** | | The spectral tool shall be controllable via commands within a Jupyter notebook cell or through a Python command line. | | | X | | **Y** |
| DL_L3_SPEC_1002.03 | | The spectral tool shall be controllable through a Data Lab GUI supplied interface within the Jupyter environment. | | | X | | N |
| **DL_L3_SPEC_1002.04** | | The spectral tool shall accept at least the data formats described by DESI and SDSS within [DESI: http://desidatamodel.readthedocs.io/en/latest/DESI_SPECTRO_REDUX/SPECPROD/spectra-NSIDE/PIXGROUP/PIXNUM/spectra-NSIDE-PIXNUM.html ; SDSS: https://data.sdss.org/datamodel/files/BOSS_SPECTRO_REDUX/RUN2D/PLATE4/spPlate.html]. | | X | | | **Y** |
| DL_L3_SPEC_1003 | Dynamic visualization of spectra | | | | | | |
| **DL_L3_SPEC_1003.01** | | The spectral tool shall pan and zoom spectra on command, including a reset button. | | | X | | **Y** |
| DL_L3_SPEC_1003.02 | | The spectral tool shall provide a zoom box automatically. | | | X | | N |
| DL_L3_SPEC_1003.03 | | The spectral tool should offer the ability to zoom in on several predefined spectral windows at once, in the form of multiple zoom boxes, on command. | Not a requirement | | | | |
| **DL_L3_SPEC_1003.04** | | The spectral tool shall overlay spectra signal components (1D spectra, sky, noise) on command. | | | X | | **Y** |
| DL_L3_SPEC_1003.05 | | The spectral tool shall display each spectra signal component (1D spectra, sky, noise) individually on command. | | | X | | N |
| **DL_L3_SPEC_1003.06** | | The spectral tool shall display each spectrum, template, and lines in observed or rest-frame wavelengths on command. | | | X | | **Y** |
| **DL_L3_SPEC_1003.07** | | The spectral tool shall reset plots to original (unzoomed, unedited) format on command. | | | X | | **Y** |
| DL_L3_SPEC_1004 | Template fitting + quality check | | | | | | |
| **DL_L3_SPEC_1004.01** | | The spectral tool shall overlay the DESI calculated redshift template (as defined by the DESI pipeline) on command. | | X | | | **Y** |
| DL_L3_SPEC_1004.02 | | The spectral tool shall overlay a user-defined redshift among existing templates on command. | | X | | | N |
| DL_L3_SPEC_1004.03 | | The spectral tool shall overlay a user-uploaded template on command. | | | X | | N |
| DL_L3_SPEC_1004.04 | | The spectral tools shall calculate a X^2 or other measurement of template fit quality. | | | X | | N |
| DL_L3_SPEC_1004.05 | | The spectral tool shall display the fitness quality value automatically. | | | X | | N |
| DL_L3_SPEC_1004.06 | | The spectral tool shall display the residual on command. | | X | | | N |
| DL_L3_SPEC_1005 | Plot locations of emission/absorption lines | | | | | | |
| **DL_L3_SPEC_1005.01** | | The spectral tool shall overlay expected locations of emission / absorption lines on command in all relevant display panels, if a redshift is available (assume redshift equals zero if none available). | | X | | | **Y** |
| DL_L3_SPEC_1005.02 | | The spectral tool shall contain a library of emission / absorption lines for user selection in the overlay process. | | X | | | N |
| DL_L3_SPEC_1005.03 | | The user shall be able to specify emission / absorption lines for display on command. | | X | | | N |
| DL_L3_SPEC_1006 | Contextual viewer | | | | | | |
| DL_L3_SPEC_1006.01 | | The spectral tool shall display a photometric image of the corresponding object to the spectra displayed. | | | X | | N |
| DL_L3_SPEC_1006.02 | | The contextual viewer should anticipate further applications of contextual capabilities (e.g. displaying a spectral energy distribution, 2D spectrum with trace, etc.). | Not a requirement | | | | |
| DL_L3_SPEC_1006.03 | | The spectral tool should give users the option to search around the displayed object within all Data Lab holdings. | Not a requirement | | | | |
| DL_L3_SPEC_1007 | Data access + saving | | | | | | |
| **DL_L3_SPEC_1007.01** | | The spectral tool shall support default units of wavelength and flux. | | X | | | **Y** |
| DL_L3_SPEC_1007.02 | | The spectral tool should support user-defined units for the x and y axes. | | X | | | N |
| DL_L3_SPEC_1007.03 | | The spectral tool shall allow users to access spectral data through a Data Lab file access service / frontend. | | X | | | N |
| **DL_L3_SPEC_1007.04** | | The Data Lab file access service / frontend shall query a spectroscopic table within a Data Lab database (e.g. queries according to ra, dec, exposure information, target class, etc.). | | | X | | **Y** |
| DL_L3_SPEC_1007.05 | | The spectral tool shall handle authentication protocols provided by the Data Lab file access service / frontend, as needed to respect proprietary data access. | | X | | | N |
| **DL_L3_SPEC_1007.06** | | The Data Lab file access service / frontend shall cache the queried spectra so that they are displayed according to the display requirements. | | X | | | **Y** |
| DL_L3_SPEC_1007.07 | | The spectral tool should allow users to access spectral data by uploading files from their local computer. | | X | | | N |
| **DL_L3_SPEC_1007.08** | | The spectral tool shall allow users to download FITS files of the displayed spectrum, including a template on command. | | X | | | **Y** |
| DL_L3_SPEC_1007.09 | | The spectral tool shall allow users to save the displayed spectrum and template within the Data Lab VOSpace on command. | | X | | | N |
| DL_L3_SPEC_1007.10 | | The spectral tool shall save plots (image files) on command. | | X | | | N |
| DL_L3_SPEC_1007.11 | | The spectral analysis steps conducted within the spectral tool shall be logged, and the log shall be saved on command. | | X | | | N |
| DL_L3_SPEC_1007.12 | | The spectral tool should save the final state of the tool following a session on command. | Not a requirement | | | | |
| DL_L3_SPEC_1008 | Catalog information + annotations | | | | | | |
| **DL_L3_SPEC_1008.01** | | Spectral tool shall display available catalog information such as spectroscopic classification and sub-class, redshift, redshift uncertainty, ZWARN flags, TARGETID and targeting bits (DESI TARGET, BGS TARGET, MWS TARGET), exposure number, fiber number, imaging data such as magnitudes, profiles, etc. automatically. | | | X | | **Y** |
| DL_L3_SPEC_1008.02 | | Upon saving a file, the spectral tool shall allow users to annotate the displayed spectra (with a flag or comment) and store annotations within the saved file on command. | | X | | | N |
| DL_L3_SPEC_1009 | Temporal analysis of spectra | | | | | | |
| **DL_L3_SPEC_1009.01** | | The spectral tool should plot the co-added spectra from multiple exposures of a particular object if available by default. | | | X | | **Y** |
| DL_L3_SPEC_1009.02 | | The spectral tool should provide information from individual exposures of a particular object on command. | | | X | | N |
| DL_L3_SPEC_1010 | Perform functions on spectra | | | | | | |
| DL_L3_SPEC_1010.01 | | The spectral tool should perform spectral addition and subtraction in flux space on command. | | X | | | N |
| **DL_L3_SPEC_1010.02** | | The spectral tool shall display a smoothed spectrum using a user-specified smoothing level and common smoothing algorithms on command, including an option to return to the unsmoothed spectrum. | | X | | | **Y** |

**Table 6. Spectral Tool Requirements Matrix**